\documentclass[aps,prb,floats,twocolumn,floatfix,showpacs]{revtex4}
\usepackage{epsf,graphicx}
\usepackage{amssymb}
\usepackage{amsmath}
\usepackage{eepic}
\usepackage[dvips]{color}
\newlength{\piclen}
\begin{document}

\title{Anomalous Hall effect in heavy electron materials}

\author{Yi-feng Yang}
\affiliation{Beijing National Laboratory for Condensed Matter Physics and \\
Institute of Physics, Chinese Academy of Sciences, Beijing 100190, China}
\date{\today}

\begin{abstract}
We propose a new empirical formula for the anomalous Hall effect in heavy electron materials based on a phenomenological two-fluid description of the $f$-electron states. The new formula incorporates two previous theories proposed by Fert and Levy in 1987 and Kontani {\em et al.} in the early 1990s and takes into account both incoherent and coherent skew scatterings from local and itinerant $f$-electrons. We perform experimental analysis in several heavy electron compounds and show that the new formula provides a consistent description of the evolution of the Hall coefficient in the whole temperature range down to only a few Kelvin. 
\end{abstract}

\pacs{71.27.+a, 72.10.-d, 72.15.Gd}
% 71.27.+a  Strongly correlated electron systems; heavy fermions
% 71.10.Fd  Lattice fermion models (Hubbard model, etc.)
% 71.30.+h  Metal-insulator transitions and other electronic transitions
% 71.20.Eh  Rare earth metals and alloys
% 72.10.-d  Theory of electronic transport; scattering mechanisms
% 72.15.Gd Galvanomagnetic and other magnetotransport effects
% 74.25.Ha  Magnetic properties 
% 74.25.Nf   Response to electromagnetic fields (nuclear magnetic resonance,etc.) 
% 74.50.+r   Tunneling phenomena; point contacts, weak links, Josephson effects
% 74.70.Tx   Heavy-fermion superconductors
% 75.20.Hr   Local moment in compounds and alloys; Kondo effect, valence fluctuations, heavy fermions
% 75.30.Mb  Valence fluctuation, Kondo lattice, and heavy-fermion phenomena 
% 75.47.Gk   Colossal magnetoresistance
% 76.60.-k     Nuclear magnetic resonance and relaxation
% 76.60.Cq   Chemical and Knight shifts 

\maketitle

\section{Introduction}
The anomalous Hall effect has attracted much interest in recent years due to its topological origin.\cite{Nagaosa2010} In general, the measured Hall coefficient $R_H$ includes two terms, $R_H=R_0+R_s$, where $R_0$ is the ordinary Hall coefficient and $R_s$ is the extra-ordinary or anomalous Hall coefficient. The microscopic origin of $R_s$ has proved quite complicated. Three distinct contributions -- intrinsic,\cite{Karplus1954} skew scattering \cite{Smit1955} and side-jump \cite{Berger1964} -- have been identified; each of them has an individual scaling, 
\begin{equation}
R_s\propto\rho^\alpha M_z/H,
\end{equation}
with respect to the longitudinal resistivity $\rho$. Here $M_z$ is the magnetization and $H$ is the magnetic field along the z-axis. In ferromagnetic conductors, a simple summation of the three terms yields an empirical formula that explains a large amount of experimental data.

In heavy electron materials, however, a satisfactory formula has not been achieved despite a number of theoretical proposals in the past three decades.\cite{Coleman1985, Fert1987, Kohno1990, Yamada1993, Kontani1994} The most prevailing theory nowadays was developed by Fert and Levy in 1987.\cite{Fert1987} They considered incoherent skew scattering of conduction electrons by independent $f$-moments and predicted,
\begin{equation}
R_s=r_l\rho\chi,
\label{Eq:RH1}
\end{equation}
where $\chi$ is the magnetic susceptibility and $r_l$ is a constant. Their theory has been verified in CeAl$_3$, CeCu$_2$Si$_2$ and most other materials in the high temperature regime but fails when coherence sets in. In the caged compound Ce$_3$Rh$_4$Sn$_{13}$, in which no lattice coherence is observed, the scaling persists down to the lowest measured temperature.\cite{Kohler2007} In most nonmagnetic Ce- and U-compounds such as CeRu$_2$Si$_2$, CeNi, CeCu$_6$, UPt$_3$ and UAl$_2$, a different scaling, 
\begin{equation}
R_s\propto\rho^2\chi,
\end{equation}
has been observed at very low temperatures and explained by Kontani {\em et al.} in the early 1990s as the coherent skew scattering of $f$-electrons.\cite{Kohno1990, Yamada1993, Kontani1994} In the nonmagnetic compound Ce$_2$CoIn$_8$, both scalings seem to apply in their respective high or low temperature regime.\cite{Chen2003} 

But unlike in ferromagnetic conductors, a direct summation of the two contributions fails to explain the experimental result in the intermediate temperature regime. The theory of Kontani {\em et al.} \cite{Kontani1994} extrapolates to a quite different scaling, 
\begin{equation}
 R_s=r_h\chi,
 \label{Eq:RH2}
 \end{equation}
where $r_h$ is a constant, and fails to describe localized $f$-moments at high temperatures in most heavy electron materials. A proper interpolation is hence required as the two formulas deal with different aspects of the $f$-electron character. Such a combination is expected to be, if possible at all, highly nontrivial, which requires detailed knowledge about the incoherent and coherent behaviors of the $f$-electrons. 

Unfortunately, we still do not have a microscopic theory that allows us to treat equally the incoherent and coherent behaviors of heavy electrons. Nonetheless, a quantitative measure of the dual feature of the $f$-electron states is now available thanks to the recent development of a phenomenological two-fluid framework.\cite{Yang2011} In this work, we take the two-fluid model as a guide and propose a new empirical formula for describing the anomalous Hall effect in heavy electron materials. We then perform data analysis in several compounds and present experimental evidence that provides unambiguous support for our formula.

\begin{figure}[t]
\centerline{{\includegraphics[width=.51\textwidth]{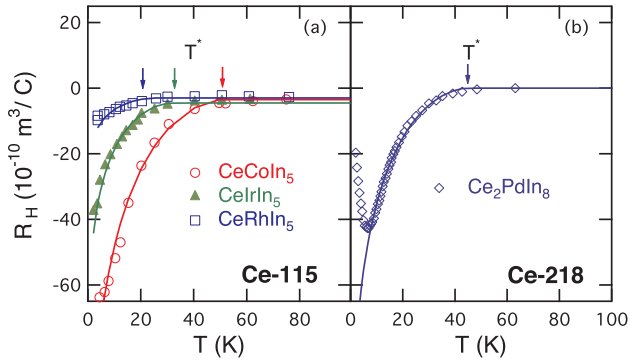}}}
\caption{
{(Color online) Hall coefficient of Ce-115 and Ce-218 materials\cite{Hundley2004,Gnida2012}. The incoherent skew scattering contribution is completely suppressed and the temperature dependence of the Hall coefficient does not follow any combination of $\rho^\alpha\chi$. Instead, it scales with the susceptibility of the itinerant Kondo liquid (solid lines).\cite{Yang2008}}
\label{Fig:Ce115}}
\end{figure}

\section{Model}

The anomalous Hall effect in Ce-115 (Ce$M$In$_5$, $M=$ Co, Ir, Rh) provides the first insight into this complicated problem. Fig. \ref{Fig:Ce115}(a) reproduces the experimental data of all three Ce-115 materials.\cite{Hundley2004,Nakajima2007} Their constant behavior at high temperatures indicates that the incoherent skew scattering contribution predicted by Fert and Levy is negligible small, namely $r_l\approx0$, and can thus be safely disregarded. Interestingly, the Hall coefficient develops a strong temperature dependence at low temperatures, accompanying the onset of coherence below $T^*\approx$20, 30 and 50 K, respectively. The temperature evolution of the Hall coefficient therefore reflects the contribution of coherent $f$-electrons. It turns out that the Hall coefficient $R_H$ cannot fit to any combination of $\rho^\alpha\chi$. Instead, within the framework of the two-fluid model, it is found to scale with a universal fraction of the magnetic susceptibility, namely the susceptibility of an itinerant heavy electron Kondo liquid,\cite{Yang2008, Yang2012} 
\begin{equation}
R_s\propto\chi_h=\chi_0\left(1-\frac{T}{T^*}\right)^{\eta}\left(1+\ln\frac{T^*}{T}\right),
\label{Eq:chih}
\end{equation}
where $\eta\approx 3/2$ and $\chi_0$ is independent of temperature. 

Quite recently, similar scaling has also been observed in Ce$_2$PdIn$_8$ with an almost constant Hall coefficient at high temperatures [Fig. \ref{Fig:Ce115}(b)].\cite{Gnida2012} Such universal $\chi_h$-scaling seems to support the prediction of Kontani {\em et al.} at high temperatures,\cite{Kontani1994} but is only subject to its own magnetization given by the heavy electrons. We hence expect that there are two seemingly separate contributions in the anomalous Hall effect related to the local and itinerant $f$-electrons, respectively. 

To see this, we consider a periodic Anderson lattice. The optical conductivity is given by the current-current correlation function,
\begin{equation}
\sigma_{\alpha\beta}=\lim_{\omega\rightarrow 0}\frac{1}{\hbar\omega}\int^\infty_0dt e^{i\omega t}\langle\left[J_\alpha(t), J_\beta(0)\right]\rangle
\end{equation}
where $\hbar$ is the Planck constant and $\alpha,\beta$ denote the $x,y$-directions. The electrical current density has two components, $J_\alpha=J_\alpha^c+J_\alpha^f$, from conduction electrons and $f$-electrons, respectively. The optical conductivity hence contains three terms
\begin{equation}
\sigma_{\alpha\beta}=\sigma_{\alpha\beta}^c+\sigma_{\alpha\beta}^f+\sigma_{\alpha\beta}^{cf},
\end{equation}
in which $\sigma_{\alpha\beta}^{c/f}$ are determined by the correlation functions of $J_\alpha^{c/f}$ for conduction and $f$-electrons separately and $\sigma_{\alpha\beta}^{cf}$ is the mixed term, which is nonzero due to hybridization.

Since $f$-electrons are well localized at high temperatures and collective hybridization develops gradually with lowering temperature, $\sigma_{\alpha\beta}^c$ is dominated by incoherent scattering inherited from high temperature local $f$-moments. On the other hand, $\sigma_{\alpha\beta}^f$ and $\sigma_{\alpha\beta}^{cf}$ only emerge as a result of collective hybridization. We can therefore recombine the three terms and get, in approximation, $\sigma_{\alpha\beta}^l=\sigma_{\alpha\beta}^c$ and $\sigma_{\alpha\beta}^h=\sigma_{\alpha\beta}^{cf}+\sigma_{\alpha\beta}^f$, with each representing either incoherent or coherent contributions separately. Without going into detail and trying to solve the complicated many-body problem, we restrict ourselves to this lowest order approximation and  seek a simple formula that can be used in experimental analysis. The Hall coefficient is then given by
\begin{equation}
R_s=\rho^2\sigma_{xy}/H=\left(\frac{\sigma_{xx}^l}{\sigma_{xx}}\right)^2R_s^l+\left(\frac{\sigma^h_{xx}}{\sigma_{xx}}\right)^2R_s^h,
\end{equation}
in which $R_s^{l}$ can be approximated only from scattering between conduction electrons and residual localized $f$-moments. Following Fert and Levy, we have $R_s^l\approx r_l\rho_l\chi_l$, where $\rho_l=1/\sigma_{xx}^l$ is the incoherent contribution to the longitudinal resistivity and $\chi_l$ is the magnetic susceptibility of the residual unhybridized $f$-moments. For the prefactor $\sigma_{xx}^l/\sigma_{xx}$, it has been shown experimentally \cite{Nakatsuji2004} that for the whole temperature range down to somewhere above the Fermi liquid temperature, the longitudinal resistivity is dominated by incoherent scattering between conduction electrons and localized $f$-moments so that $\rho\approx\rho_l$, or $\sigma_{xx}^l/\sigma_{xx}\approx1$. This approximation becomes exact for $T>T^*$, where all $f$-electrons behave as fully localized magnetic moments. At very low temperatures where only coherent heavy electrons exist, the approximation fails but the formula still holds because of the suppression of $R_s^l$ with $\chi_l=0$. 

Similarly, the second term only starts to contribute following the emergence of heavy $f$-electrons and can therefore be treated purely as a coherent effect due to collective hybridization. It was not obtained in the theory of Fert and Levy but only derived later in a coherent treatment of the periodic Anderson model by Kontani {\em et al.}.\cite{Kontani1994} They find $R_s^h\propto \gamma_h^2\chi_h/(E_f^2+\gamma_h^{2})$, where $E_f$, typically of the order of $T^*$, is the renormalized $f$-electron energy relative to the Fermi energy, and $\gamma_h$ is the imaginary part of the $f$-electron self-energy, varying from a few meV at zero temperature to $\sim 100\,$meV near $T^*$.\cite{Shim2007,Yang2009,Schoenes1987} This leads to the scaling $R_s^h\propto\chi_h$ at high temperatures and $R_s^h\propto\rho^2_h\chi_h=\rho^2\chi$ at extremely low temperatures in the Fermi liquid regime, where $\rho=\rho_h=1/\sigma_{xx}^h\propto \gamma_h$ and $\chi=\chi_h$. For $r_l=0$, the $\chi_h$-scaling in Ce-115 therefore suggests that the prefactor $\sigma_{xx}^h/\sigma_{xx}$ is only weakly temperature-dependent, at least in the intermediate temperature range, before it approaches unity in the Fermi liquid regime. This is further supported in the dynamical mean-field theory (DMFT) calculations.\cite{Choi2012} A crossover from $\rho^0$ to $\rho^2$-scaling is hence expected at a certain temperature and may be responsible for the minimum in the Hall coefficient of Ce$_2$PdIn$_8$. 

Although we do not have a microscopic theory for the exact behavior of the optical conductivity of an Anderson lattice, the above argument at least suggests an approximate two-component description of the Hall coefficient that could be easily applied to experimental analysis,
\begin{equation}
R_H\approx R_0+r_l\rho\chi_l+r_h\chi_h,
\label{Eq:Rh}
\end{equation}
where $R_0$, $r_l$ and $r_h$ are all assumed to be constant. For $T>T^*$, we have $\chi_h=0$ and the formula reduces to $R_H=R_0+r_l\rho\chi$; for $T<T_{FL}$, we have instead $\chi_l=0$ and $R_H=R_0+\tilde{r}_h\rho^2\chi$, where $\tilde{r}_h$ is another constant prefactor different from $r_h$. The quite unusual form of $R_H$ reflects the fundamental difference of heavy electron materials from ferromagnetic conductors in that, with lowering temperature, localized $f$-moments gradually dissolve into the Kondo lattice and may not be treated as a static spin polarized background. This is in fact a general feature of strongly correlated $d$-electron and $f$-electron systems \cite{Barzykin2009} and has not been taken into account in all previous proposals.\cite{Nagaosa2010} In this sense, Eq. (\ref{Eq:Rh}) provides not only a new empirical formula for experimental analysis, but also a completely new possibility for future theoretical investigation. 

\begin{figure}[t]
\centerline{{\includegraphics[width=.48\textwidth]{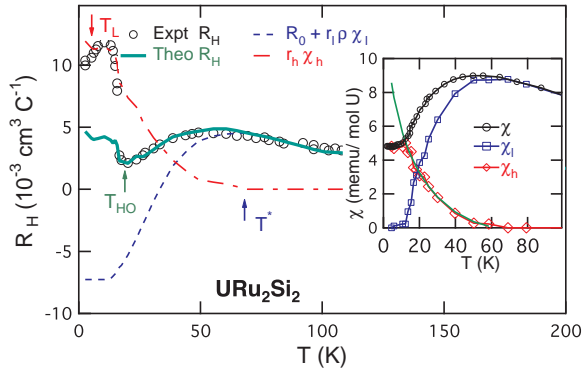}}}
\caption{
{(Color online) Analysis of the Hall coefficient of URu$_2$Si$_2$.\cite{Schoenes1987} The ordinary Hall coefficient is $R_0=-7.27\times10^{-3}\,$cm$^{3}$C$^{-1}$. We have $T^*=65\,$K and $T_L=10\,$K. Using the new formula, we find a good fit in the whole temperature range above $T_{HO}$. The inset plots the two components of the susceptibility derived from the Knight shift analysis.\cite{Yang2012,Bernal2000,Curro2012} The derived $\chi_h$ follows exactly the scaling of the Kondo liquid (solid line).}
\label{Fig:URS}}
\end{figure}

\section{Experimental Analysis}

We apply the above formula to experiment. As will be shown below, our formula tracks the whole temperature evolution of the Hall coefficient down to only a few Kelvin, below which either a new phase emerges or a crossover shows up entering a Fermi liquid ground state. The first and obvious prediction of the new formula in the limit $r_l\approx0$, namely the scaling $R_H=R_0+r_h\chi_h$, after a temperature-independent Hall coefficient in the high temperature incoherent regime, has been confirmed in La-doped CeCoIn$_5$\cite{Yang2008} and all three Ce-115 compounds under pressure\cite{Hundley2004,Nakajima2007} and will not be repeated here. In Fig.~\ref{Fig:Ce115}(b), the very recent experiment on Ce$_2$PdIn$_8$ once again verifies the $\chi_h$-scaling.\cite{Gnida2012}

We need to consider materials with a non-negligible incoherent contribution. The key point is to split the total magnetic susceptibility into two components, $\chi_l$ and $\chi_h$, so that the formula can be applied to experiment in a straightforward manner. URu$_2$Si$_2$ provides a typical example, in which the Knight shift experiment \cite{Bernal2000,Curro2012} allows us to determine $\chi_l$ and $\chi_h$ unambiguously. The two-fluid description of the magnetic susceptibility and the Knight shift \cite{Nakatsuji2004,Curro2004,Yang2008} reads, $\chi=\chi_l+\chi_h$ and $K=K_0+A\chi_l+B\chi_h$, where $A$ and $B$ are the hyperfine couplings. At high temperatures, $f$-electrons are well localized so that $K=K_0+A\chi$ for $T>T^*\approx65\,$K; while at very low temperatures deep within the hidden order phase,\cite{Mydosh2011} all 5$f$-electrons become itinerant and we have $K=K_0+B\chi$ for $T<T_L\approx 10\,$K. These determine the values of $K_0$, $A$ and $B$ without arbitrariness. The two components $\chi_l$ and $\chi_h$ for $T_L<T<T^*$ are then immediately derived by using $\chi_l=(K-K_0-B\chi)/(A-B)$ and $\chi_h=(K-K_0-A\chi)/(B-A)$.\cite{Yang2012,Curro2012} The results are plotted in the inset of Fig.~\ref{Fig:URS}. As we can see, the partial susceptibility $\chi_h$ (diamonds) falls exactly upon the scaling formula (solid line) of the Kondo liquid predicted in the two-fluid framework.\cite{Yang2008}

\begin{figure}[t]
\centerline{{\includegraphics[width=.5\textwidth]{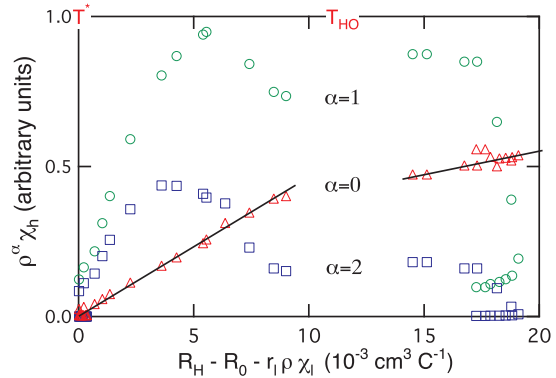}}}
\caption{
{(Color online) Comparison between the subtracted coherent contribution $R_H-R_0-r_l\rho\chi_l$ and $\rho^\alpha\chi_h$ for $\alpha=\,$0, 1, 2 in URu$_2$Si$_2$. Only for $\alpha=0$, is an overall proportionality found in the whole temperature range between $T^*$ and $T_{HO}$.}
\label{Fig:URS2}}
\end{figure}

We now turn back to the Hall measurements on URu$_2$Si$_2$.\cite{Schoenes1987} In the high temperature regime $T>T^*$, all $f$-electrons are localized and, by using the experimental data for the magnetic resistivity,\cite{Schoenes1987} the Hall coefficient is found to scale very well with the prediction of Fert and Levy up to the highest measured temperature of about $300\,$K,
\begin{equation}
R_H=R_0+r_l\rho \chi,
\label{Eq:Rl}
\end{equation}
which determines the values of $R_0$ and $r_l$. We can then subtract the true incoherent contribution $r_l\rho\chi_l$ to obtain the possible forms of the coherent contribution. Fig. \ref{Fig:URS2} compares $R_H-R_0-r_l\rho\chi_l$ and $\rho^\alpha\chi_h$ calculated from experimental data for different values of $\alpha$. Only for $\alpha=0$ do we find an overall agreement for temperatures between $T_{HO}$ and $T^*$. This provides an unambiguous support to Eq.~(\ref{Eq:Rh}) without any adjustment of parameters. In Fig.~\ref{Fig:URS}, an excellent overall fit is seen in the whole temperature range above $T_{HO}$. For $T<T_{HO}$, different values of $R_0$ and $r_h$ are required to fit the data, indicating a Fermi surface change across the hidden order transition.\cite{Mydosh2011}

The above analysis can be easily extended to other materials. If no additional experimental information is available for the separation of $\chi_l$ and $\chi_h$, a less rigorous method may be applied by using {\em a priori} the scaling formula in Eq.~(\ref{Eq:chih}).\cite{Yang2008} Once again, $R_0$ and $r_l$ can be determined from the high temperature fit. We then analyze the data in the intermediate temperature regime by using
\begin{equation}
R_H=R_0+r_l\rho(\chi-\chi_h)+r_h\chi_h.
\label{Eq:Rh2}
\end{equation}

\begin{figure}[t]
\centerline{{\includegraphics[width=.48\textwidth]{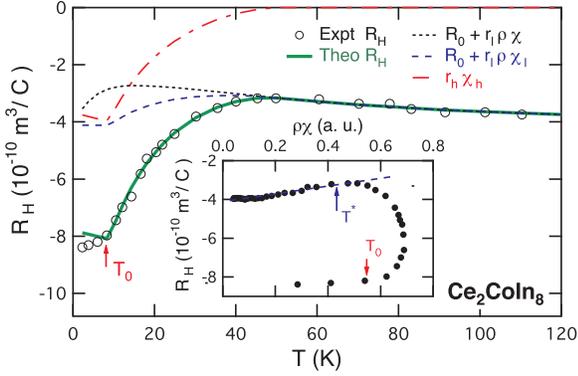}}}
\caption{
{(Color online) Analysis of the Hall coefficient of Ce$_2$CoIn$_8$.\cite{Chen2003} The parameters are $T^*=50\,$K, $R_0=-4.12\times10^{-10}\,$m$^3$/C, and $\chi_0=5.0\times10^{-3}\,$emu/mol-Ce. For comparison, the dotted line shows the prediction of Fert and Levy with a constant $r_l$. The inset plots $R_H$ vs $\rho\chi$. }
\label{Fig:Ce218}}
\end{figure}

As an example, Fig.~\ref{Fig:Ce218} shows the fit to the Hall data in Ce$_2$CoIn$_8$.\cite{Chen2003} The inset compares $R_H$ with $\rho\chi$ calculated from experiment. The high temperature behavior of the Hall coefficient is well described by incoherent skew scattering. Below $T^*\approx50\,$K, the prediction of Fert and Levy in Eq. (\ref{Eq:Rl}) (dotted line) deviates severely from experiment. On the other hand, using the new formula and taking into account the coherent contribution, we obtain a good fit all the way down to $T_0\approx10\,$K, below which a $\rho^2$-dependence was claimed experimentally.\cite{Chen2003} For comparison, Fig.~\ref{Fig:Ce218} also plots the two contributions separately  and the coherent skew scattering is seen to have a major contribution (dash-dotted line) in Ce$_2$CoIn$_8$. In Fig.~\ref{Fig:Ce115}(b), the incoherent skew scattering is even more suppressed in the isostructural compound Ce$_2$PdIn$_8$,\cite{Gnida2012} giving rise to a similar $\chi_h$-scaling that was discussed previously in Ce-115 compounds. 

In contrast, the anomalous Hall effect in YbRh$_2$Si$_2$ \cite{Paschen2005} is dominated by incoherent skew scattering. While its coherence temperature is $\sim70\,$K,\cite{Yang2008b,Mo2012} $R_H$ follows roughly the prediction of Eq. (\ref{Eq:Rl}) for all temperatures above $\sim7\,$K with only a slight change of slope ($r_l$) at $\sim 90\,$K,\cite{Paschen2005} possibly due to the crystal field effect.\cite{Ernst2011} The suppression of the itinerant term may be understood from accidental cancellation of positive and negative contributions from different sheets of the Fermi surface.\cite{Norman2005,Wigger2007} In particular, YbRh$_2$Si$_2$ also exhibits little sign of the Knight shift anomaly,\cite{Ishida2002} unlike most other heavy electron materials, and its origin needs to be further explored. Because of this suppression, by applying an external pressure or magnetic field, a clear crossover in the Hall measurement should show up when all localized $f$-moments change their character and become itinerant, accompanied by a reconstruction of the Fermi surface.\cite{Yang2012} This has been observed in YbRh$_2$Si$_2$.\cite{Paschen2004} The signature is expected to be less prominent in other materials.

All together, while Ce-115, Ce$_2$PdIn$_8$ and YbRh$_2$Si$_2$ represent rare extremes with almost completely suppressed incoherent or coherent contributions, most compounds are like Ce$_2$CoIn$_8$ and URu$_2$Si$_2$ and exhibit the two-component physics. Our analysis confirms the proposed formula in Eq.~(\ref{Eq:Rh}).

\section{Discussions}

Our results suggest a new categorization of the rich variety of the Hall behaviors in heavy electron materials: 

(i) Above the characteristic temperature $T^*$ determined by the onset of coherence,\cite{Yang2008b} there are only localized $f$-moments and their incoherent skew scattering gives $R_H=R_0+r_l\rho\chi$.\cite{Fert1987}

(ii) In the Fermi liquid regime, where all $f$-electrons are itinerant, the anomalous Hall effect is given by coherent scattering alone so that $R_H=R_0+\tilde{r}_h\rho^2\chi$. The temperature region of this $\rho^2$-scaling depends on the details of hybridization.\cite{Kontani1994}

(iii) In between, the normal state $f$-electrons are both dynamically itinerant and localized so that both incoherent and coherent scatterings contribute and give rise to the unusual formula, $R_H\approx R_0+r_l\rho\chi_l+r_h\chi_h$. 

(iv) In special cases where incoherent skew scattering is completely suppressed ($r_l=0$), one finds a universal $\chi_h$-scaling.\cite{Yang2008} Ce-115 \cite{Hundley2004,Nakajima2007} and Ce$_2$PdIn$_8$ \cite{Gnida2012} fall into this category, but the origin of the suppression has not been explained. In the theory of Fert and Levy, the incoherent contribution originates from the interference between the $f$ and $d$ partial waves, with $r_l\propto\sin\delta_2\cos\delta_2$, where $\delta_2$ is the phase shift of the $d$-scattering channel.\cite{Fert1987} A detailed study of the distinction between Ce$_2$CoIn$_8$ and Ce$_2$PdIn$_8$ may be able to determine $\delta_2$ and help resolve this issue. 

The specific conditions that control the relative importance of $R_s^l$ and $R_s^h$ in the two-fluid regime are not known. Theoretically, although we were partly motivated by the theory of Kontani {\em et al.},\cite{Kontani1994} its validity for heavy electron materials remains unclear. A notable prediction of the theory is that the sign of $r_h$ depends on the location of the renormalized $f$-electron energy around the Fermi energy so that Ce-compounds would have a positive $r_h$ and Yb-compounds would have a negative $r_h$. Instead, we find negative values of $R_s^h$ and $R_0$ in Ce-115, Ce$_2$CoIn$_8$ and Ce$_2$PdIn$_8$ and positive values of $R_s^h$ and $R_0$ in YbRh$_2$Si$_2$, in line with the electron or hole nature of their charge carriers. A thorough understanding of the new formula may hence require a microscopic theory of the two-fluid physics that is not yet available. Nevertheless, the fact that $\chi_h$-scaling is in good agreement with the quasiparticle density of states calculated by DMFT \cite{Shim2007} seems to suggest that DMFT may provide an explanation to our empirical formula by separating the contribution of coherent quasiparticles and that of incoherent skew scattering due to incomplete screening of the $f$-moments at finite temperatures. We will leave this for future work.

In conclusion, we propose a new empirical formula for the anomalous Hall effect in heavy electron materials. The formula makes it possible for better data analysis and allows for the first time a consistent interpretation of the Hall experiment over a broad temperature range down to only a few Kelvin. It unifies the various scalings observed in different temperature regimes in experiment and provides a new basis for developing an improved theory incorporating previous proposals and the two-fluid physics. The identification of incoherent and coherent contributions opens a new avenue for future numerical investigations. A similar phenomenon may also be found in other physical properties such as the spin Hall effect.

\section{Acknowledgements}

The author thanks D. Pines for the stimulating discussion. This work is supported by NSF-China (Grant No. 11174339) and the Chinese Academy of Sciences.

\end{document}